\documentclass[twocolumn,amsmath,amssymb,aps,prl]{revtex4}

\usepackage{graphicx}

\begin{document}
\pagestyle{empty}
\preprint{MANUSCRIPT}

\title{Measuring entanglement using quantum quenches}

\author{John Cardy}
\affiliation{Rudolf Peierls Centre for Theoretical Physics, Oxford
University, 1 Keble Road, Oxford, OX1 3NP, United Kingdom}
\affiliation{All Souls College, Oxford, United Kingdom}
\affiliation{Kavli Institute for Theoretical Physics, Santa
Barbara}

\date{March 7 2011}

\begin{abstract}
We show that block entanglement entropies in one-dimensional
systems close to a quantum critical point can in principle be
measured in terms of the population of low-lying energy levels
following a certain type of local quantum quench.
\end{abstract}

\maketitle

Entanglement, and its quantification through entanglement entropy,
have become increasingly important tools in the study of quantum
many-body systems. The fact that the entanglement entropy of a
region of linear size $\ell$ in the ground state of a system with
short-range interactions in $d$ dimensions grows like $\ell^{d-1}$
\cite{arealaw} (up to possible logarithms), as compared with a
typical state where it is extensive, both explains the success of
modern numerical methods and informs their further development.
Moreover it gives a basis-independent way of detecting quantum
critical behavior and topological phases.

Briefly, the entanglement entropy is defined as follows: given a
bipartition of the Hilbert space ${\cal H}={\cal H}_A\otimes{\cal
H}_B$ (which usually corresponds to the degrees of freedom lying
in mutually exclusive regions $A$ and $B$ of $d$-dimensional
space), the reduced density matrix of (say) $A$ is given by
$\rho_A=\text{Tr}_{{\cal H}_B}\,|0\rangle\langle0|$. The R\'enyi
entropies are then given by
$S_A^{(n)}=(1-n)^{-1}\log\text{Tr}_{{\cal H}_A}\,\rho_A^n$. The
von Neumann entropy $-\text{Tr}_{{\cal H}_A}\,\rho_A\log\rho_A$ is
formally the limit as $n\to1$. For different $n$ these encode the
entanglement spectrum of $\rho_A$: if we make a Schmidt
decomposition
\begin{equation}\label{eq:schmidt}
|0\rangle=\sum_kc_k|k\rangle_A\,|k\rangle_B
\end{equation}
then the non-zero eigenvalues of $\rho_A$ are given by $c_k^2$. On
the other hand $\text{Tr}_{{\cal H}_A}\,\rho_A^n=\sum_kc_k^{2n}$.
If the state has low entanglement then only a few eigenvalues are
appreciable and the entropies are small, while if the state is
maximally entangled the entropies are of the order of the lesser
of the dimensions of ${\cal H}_A$ and ${\cal H}_B$.

In exactly solvable models, and more generally for systems at or
near a quantum critical point, considerable progress has been made
in relating the behavior of the entanglement entropy to other
universal data of the underlying long-distance theory, typically a
quantum field theory (QFT) \cite{ccreview}. In particular, for
$d=1$, the coefficient of the `area' law (in this case the number
of boundary points between $A$ and $B$) is predicted to diverge as
$\log\xi$, where $\xi$ is the correlation length, as the critical
point is approached. At the critical point, the entropy of a
finite interval of length $\ell$ immersed in a much larger system
grows like $\log\ell$. When the low energy, long wavelength
physics is described by a conformal field theory (CFT), the
coefficient is given by the central charge, or conformal anomaly
number, $c$: the R\'enyi entropies behave as \cite{holzhey,cc1}
$S_A^{(n)}\sim(c/6)(1+n^{-1})\log\ell$. Many other predictions
have been made by now. For example, when $A$ consists of two
disjoint intervals, the entropy encodes all the data of the CFT,
which is therefore recoverable solely from properties of the
ground state wave function \cite{cct1,cct2}. These predictions,
both in $d=1$ and higher, have been well verified in numerical
investigations, and indeed the low degree of entanglement is
partly responsible for the success of DMRG and tensor network
methods \cite{cirac}.

However it seems difficult to devise a method whereby entanglement
entropy in an extended system of the type discussed above could be
measured, even in principle, in a real experiment, since it is
intrinsically non-local. While various methods have been proposed
for measuring entanglement in systems with a finite dimensional
Hilbert space \cite{HE}, their complexity increases with the
system size. What we have in mind is a conventional condensed
matter experiment whose difficulty does not in principle increase
with the system size. In some simple systems the entanglement
entropy can be indirectly recovered from an (in principle)
measurable correlation function \cite{peschel,hoyos} or the
distribution of suitably chosen observables \cite{KRS}, but this
connection is system-specific.


Various suggestions have been made as to how charge
\cite{kl,song1} or number \cite{song2} fluctuations in the
subsystem $A$, either in space or in time, may provide a
measurement of entanglement entropy. However, the generality of
these observations has been questioned \cite{fradkin}. They are
restricted to systems with a conserved current, and it is
difficult to see how these ideas could measure entanglement of
neutral degrees of freedom, or apply in cases when there is no
such conservation law. More seriously, the logarithmic behavior
which appears in these analyses can be traced to the fact that, in
one dimension, current-current correlators $\langle
J(x_1)J(x_2)\rangle$ behave like $|x_1-x_2|^{-2}$ on separations
much smaller than the size of the subsystem (but larger than the
microscopic scale, and similarly in the time domain), giving rise
to logarithms on integration over $x_1$ and $x_2$. Although the
functional form of such logarithms turns out to be identical to
that of the entanglement entropy for simple geometries, the
analysis of Ref.~\onlinecite{cct2} shows that, even in the
slightly more complicated case where $A$ consists of two disjoint
intervals, the entanglement entropy has a form which cannot be
simply expressed as an integral over correlation functions of
local fields.

The proposal in Ref.~\onlinecite{kl} is an example of a local \em
quantum quench \em\cite{local,cclocal}, whereby the hamiltonian of
a quantum system is instantaneously changed $ H\to H'$ in a local
way, so that the quantum state, which was the ground state of $
H$, now evolves according to $ H'$. In general, the real time
behavior after such a local quench is relatively simple
\cite{cclocal}: the additional energy near the quench site is
radiated away as the quasiparticles of $ H'$, moving
semi-classically at their group velocities $v_g$. These also
propagate \em changes \em in entanglement through the system,
which, in a critical 1d system, therefore grow like $\log(v_gt)$.

In this letter we propose a different type of local quantum quench
which, in principle, directly measures the R\'enyi entropies. We
show that these are simply given by the probability
$P_0=|{}_{H'}\langle0|0\rangle_H|^2$ of finding the system in the
ground state of $ H'$ after the quench. This quantity is called
the \em fidelity\em, and it has been used extensively in
characterizing quantum critical behavior (see, for example, Ref.
\onlinecite{fidelity}), but it has largely been restricted to
global, rather than local changes in the hamiltonian. (See,
however, \cite{dubail}.) It may be objected that $P_0$ is not
directly experimentally observable, even in a gapped system.
However we also show that, in systems close to a quantum critical
point described by a QFT with a linear dispersion relation, the
probability $P(E)$ of finding the system in a low-lying excited
state of energy $E$ is given by $P_0$ times a \em calculable \em
factor. In principle, $P(E)$ is measurable if the system is
coupled weakly in a known fashion to other modes whose spectrum
can be analyzed. This then gives an indirect measurement of $P_0$
and hence the R\'enyi entropy. The analog of $P(E)$ for global
quenches has been discussed in \cite{Pol,Silva}.

Let us define more precisely the local quench. In computing
R\'enyi entropies, it is convenient to imagine $n$ copies of the
original system. We now suppose that these copies can actually be
manufactured, to a tolerance to be discussed later. Consider
therefore $n$ identical copies of the system so that the full
Hilbert space is ${\cal H}=\otimes_{j=1}^n{\cal H}_j$. Initially
they are decoupled so that the hamiltonian is $H=\sum_{j=1}^nH_j$,
and each system is in its ground state (assumed unique), so the
combined system is in its ground state $|0\rangle_H=
\prod_j|0\rangle_j$. Consider the same bipartite decomposition of
each ${\cal H}_j={\cal H}_{jA}\otimes{\cal H}_{jB}$. All the
${\cal H}_{jA}$ are isomorphic. Let $\Pi_n$ be the permutation
operator (unitary on $\cal H$) which maps ${\cal H}_{jA}\to{\cal
H}_{(j+1)A}$ (mod $n$) and acts as the identity on all the ${\cal
H}_{jB}$. (Such an operator for the case $n=2$ was introduced in
Ref.~\onlinecite{hastings} where it is called a swap operator.)
Then it is well-known (see e.g. Ref.~\onlinecite{Z1} for the case
$n=2$) that using the decomposition (\ref{eq:schmidt}) and the
orthonormality of the Schmidt states,
\begin{equation}\label{eq:fid}
{}_H\langle0|\Pi_n|0\rangle_H=\sum_kc_k^{2n}=\text{Tr}_{{\cal
H}_A}\,\rho_A^n
\end{equation}
On the other hand, if we define $H'=\Pi_n^{-1}H\Pi_n$, $H'$ and
$H$ are isospectral, and $|0\rangle_{H'}=\Pi_n^{-1}|0\rangle_H$ is
the ground state of $H'$. Hence, thinking of $H\to H'$ as a
quantum quench, the modulus squared of (\ref{eq:fid}) is nothing
but $P_0=|{}_{H'}\langle0|0\rangle_H|^2$.

The point about this elementary observation is that, when $H$ has
only short-ranged interactions, and $A$ and $B$ are spatially
disjoint regions, the difference between $H$ and $H'$ is
restricted to the boundary between $A$ and $B$. Thus while the
action of $\Pi_n$ on states, expressed in particular basis, may be
complicated (see, \em e.g. \em Fig.~1 of
Ref.~\onlinecite{hastings}), its action on the combined
hamiltonian is simple. As a example, consider two copies of a
Heisenberg spin chain with nearest neighbor interactions and
hamiltonian $J\sum_l{\bf\sigma}_l\cdot{\bf\sigma}_{l+1}$. Take $A$
to be the set of sites with $l\leq0$ and $B$ those with $l\geq1$.
Then
$$
H'-H=J{\bf\sigma}^{(1)}_0\cdot{\bf\sigma}^{(2)}_1
+J{\bf\sigma}^{(2)}_0\cdot{\bf\sigma}^{(1)}_1
-J{\bf\sigma}^{(1)}_0\cdot{\bf\sigma}^{(1)}_1
-J{\bf\sigma}^{(2)}_0\cdot{\bf\sigma}^{(2)}_1
$$
\begin{figure}
\includegraphics[width=0.3\textwidth]{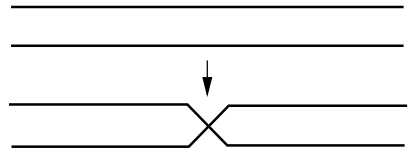}
\caption{\label{fig1}The action of a single twist operator on two
one-dimensional systems with short-range interactions.}
\end{figure}
\begin{figure}
\includegraphics[width=0.3\textwidth]{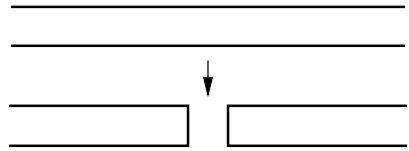}
\caption{\label{fig2}A single twist is equivalent to the closing
of a point contact when $n$ is even and the ground state is
invariant under reflection.}
\end{figure}
This is illustrated in Fig.~\ref{fig1}. (For an odd number of such
twists it is also necessary to twist the boundary conditions at
the same time. This has no effect in open systems or for periodic
boundary conditions in systems much larger than the correlation
length, but is important for finite ungapped systems with periodic
boundary conditions.) Such twist operators have been studied for
integrable spin chains in \cite{CAD}.

Note that for an infinite system or one with periodic boundary
conditions, one of the chains can be parity reversed as in
Fig.~\ref{fig2} (assuming the ground state $|0\rangle$ is
invariant under this reflection) to obtain a local quench
corresponding to the closing of a point contact coupling to two
external leads on either side. In a real experiment, of course, it
may be difficult to ensure that the new couplings added in $H'$
are precisely the same as those deleted in $H$. However, at least
in a gapped system, the results should be robust to such
imperfections as long as they are on scales smaller than the gap,
or the energy $E$ of excited states.

The operators $\Pi_n$ corresponding to a local modification of $H$
at some point $l$ in a one dimensional system are called \em twist
operators \em ${\cal T}_n(l)$ in the literature \cite{twist}. From
the above we see that the R\'enyi entanglement entropy between the
left and right halves of such a system is given in terms of the
ground state expectation value ${}_H\langle0|{\cal
T}_n(0)|0\rangle_H$. Similarly the R\'enyi entanglement entropy
between an interval of length $\ell=|l_1-l_2|$ and the rest of the
system is given by ${}_H\langle0|{\cal T}_n(l_1){\cal
T}_n(l_2)|0\rangle_H$, and so on. In all cases, these matrix
elements may be viewed equivalently as the fidelity following a
quantum quench of $H$ to $H'=\prod_k{\cal
T}_n(l_k)^\dag\,H\,\prod_k{\cal T}_n(l_k)$, or equivalently the
probability $P_0$ that the system is found in the ground state of
$H'$ when its energy is measured.

In one-dimensional systems close to a critical point, considerable
effort has gone into analyzing the behavior of correlators of
these twist operators, especially for systems with dynamical
scaling exponent $z=1$, which are described in the scaling limit
by a relativistic QFT, and at the critical point, by a CFT
\cite{ccreview}. In these cases, it follows from the early work of
Ref.~\onlinecite{holzhey} that ${\cal T}_n$ has scaling dimension
$x_n=(c/12)(n-1/n)$, where $c$ is the conformal anomaly number of
the CFT. By scaling, this means for a single twist in a system
with a finite gap $\Delta$, $P_0$ scales as $b_n'\Delta^{2x_n}$ as
$\Delta\to0$. Similarly, for an interval of length $\ell$ at the
critical points, it behaves like $b_n\ell^{-2x_n}$. The constants
$b_n$ and $b_n'$ are not universal, but their ratio is expected to
be, in units where $v=1$ \cite{doyon1}.

However, measuring the total energy of a many-body system to
accuracy $O(1)$ is unfeasible even when the spectrum is gapped. In
fact, after such a local quench, most of the excess energy goes
into states whose energy is of the order of the inverse cut-off,
or band width. This can be seen by studying the time dependence of
the energy density, given by the component $T_{00}(x,t)$ of the
energy momentum tensor, following the action of a single twist.
This is given in the Schr\"odinger picture by ${}_H\langle0|{\cal
T}_n(0)^\dag e^{iHt}T_{00}(x)e^{-iHt}{\cal T}_n(0)|0\rangle_H$. On
the other hand, in a CFT, we can write $T_{00}=T+\overline T$,
and, in imaginary time
\begin{eqnarray*}
&&{}_H\langle0|{\cal T}_n(0)^\dag e^{-H\tau'}T(x)e^{-H\tau}{\cal
T}_n(0)|0\rangle_H\\
&&=\frac{(c/24)(1-n^{-2})(\tau+\tau')^2}{(x+i\tau)^2(x-i\tau')^2}
\end{eqnarray*}
This follows from Eq.~(11) of \cite{cc1}. A similar result holds
for $\langle\overline T\rangle$. Continuing this result to
$\tau=it+\delta$ and $\tau'=-it+\delta$, where $\delta$ is a UV
cut-off of the order of the lattice spacing or inverse band width,
we see that, after the quench,
$$
\langle T_{00}(x,t)\rangle\sim \frac
c{12}(n-n^{-1})\left(\frac{\delta^2}{\big((t+x)^2+\delta^2\big)^2}
+ \{x\to-x\}\right)
$$
in units where $v=1$. The total energy $\int\langle T_{00}\rangle
dx$ diverges like $\delta^{-1}$, and, in a CFT, it is concentrated
in a region of width $O(\delta)$ near the light cone. In a more
general lattice theory we expect the energy to be carried off by
quasiparticles moving semi-classically with their appropriate
group velocities. Similar considerations apply to \em changes \em
in the entanglement entropy \cite{cclocal}.

We now argue that, despite the fact that most of the energy is
radiated in this non-universal manner, the population $P(E)$ of
the states with energies $E$ much less than the cut-off is
universal, and is directly related to the R\'enyi entropy. This
because its Laplace transform is given by correlation functions of
twist operators in \em imaginary \em time:
\begin{eqnarray*}
\int P(E)e^{-E\tau}dE&=&{}_H\langle0|e^{-(H'-H)\tau}|0\rangle_H\\
&=& {}_H\langle0|\big(\prod_k{\cal
T}_n(l_k,\tau)^\dag\big)\big(\prod_k{\cal T}_n(l_k,0)|0\rangle_H
\end{eqnarray*}
We now consider various simple cases of this.

\em Single twist in an infinite ungapped system. \em At the
critical point the two-point function $\langle{\cal
T}_n(0,\tau)^\dag{\cal T}_n(0,0)\rangle\sim b_n\tau^{-2x_n}$,
where $b_n$ is the \em same \em constant appearing in the result
for the R\'enyi entropy of an interval of length $\ell$ (in units
where $v=1$.) In this case the spectrum of $H'$ is continuous and
we see that $P(E)dE\sim \big(b_n/\Gamma(2x_n)\big)E^{2x_n-1}dE$.
Note that the probability of finding the system in a state of
energy $<E$ approaches zero as $E\to0$. This is an example of an
`orthogonality catastrophe' as was first observed in the X-ray
edge singularity \cite{Xray}. As in that problem, the probability
of occupation of low-lying states obeys a universal power law
\cite{Xraypower}.

\em Single twist in a gapped system. \em When the theory is
gapped, the 2-point function behaves asymptotically like
$|\langle{\cal T}_n\rangle|^2$, giving a probability $P_0\sim
b_n'\Delta^{2x_n}$ that the system is in the ground state. More
interesting is the occupation of low lying states. In
\cite{doyon1} it was shown that ${\cal T}_n$ couples to
multi-particle states in the continuum. As an example, the
coupling to 2-particle states with $E>2\Delta$ gives
$$
P(E)=P_0\,\sum_{k_1,k_2}\sum_{i=1}^n
\sum_{j=1}^n|F_{ij}(k_1,k_2)|^2\delta(E-E_{k_1}-E_{k_2})
$$
where $F_{ij}$ is a form factor coupling ${\cal T}_n$ to a
particle of momentum $k_1$ in copy $i$ and of momentum $k_2$ in
copy $j$. For integrable models these are calculable \cite{doyon1}
and thus measurement of $P(E)$ gives direct access to $P_0$ and
therefore the R\'enyi entropy.

\em Single twist in a finite ungapped system. \em Suppose the
initial system has finite length $L$ with open boundary
conditions. The form of the 2-point function $\langle{\cal
T}_n(x,\tau)^\dag{\cal T}_n(x,0)\rangle$ can be found by a
conformal mapping to the upper half plane. As $\tau\to\infty$ we
find $P_0=b_ng_n\left((L/\pi)\sin(\pi x/L)\right)^{-2x_n}$, where
$x$ is the distance of the twist from one end of the system, and
$\log g_n$ gives the boundary entropy. This agrees with the result
in \cite{cc1} for the R\'enyi entropy. The coupling to the excited
states can be found using the methods described in \cite{cct2}.
The leading corrections as $\tau\to\infty$ come from two
excitations each of energy $\pi x_s/L$ (where $x_s$ is a boundary
scaling dimension) propagating in two different copies of the
available $n$. We then find that
$$
P(E=2\pi x_s/L)=P_0d_n(x_s)\left((L/\pi)\sin(\pi
x/L)\right)^{4x_b}
$$
where $d_n(x_s)$ is a known function of $x_s$ \cite{cct2}.


\em Multiple twists. \em This is more  complicated still, since
for $p$ twists we need to know the $2p$-point function of twist
fields. However analytic results are available from CFT for
various limiting cases. For example take the case $p=2$, that is
an interval $A$ of length $\ell$ in an infinite system. As
discussed above, the probability $P_0$ of finding the system in
its ground state gives the R\'enyi entropy. For $\ell\ll\tau$,
that is energies $E\ll\ell^{-1}$, the methods of \cite{cct2} show
that the product of two twist operators can be written as an
infinite sum over all possible local scaling fields of the CFT on
each of the $n$ copies. The leading correction comes from when two
of these have dimension $x\not=0$ and all the rest correspond to
the identity. Thus we have
$$
P(E)dE=P_0\sum_x\tilde d_n(x)(\ell E)^{4x-1}dE+\cdots
$$
where $\tilde d_n(x)$ is once again known, the sum is over all the
bulk scaling dimensions, and the neglected terms, of order $(\ell
E)^{2(x_i+x_j+x_k)}$ are also calculable and, for $n>2$, encode
further CFT data. (For $n=2$ a closed form for $P(E)$ is available
since the 4-pt function of twist operators is related to the
partition function of the CFT on a torus \cite{cct2}.) For
$E\gg\ell^{-1}$ we find the result for single, independent twists.

Another important example of a local quench, however not related
to the R\'enyi entropy, is where two pieces of a system are joined
together to form a larger one. The real time evolution after such
a quench was studied in \cite{local,cclocal}. More recently the
fidelity was evaluated using CFT methods \cite{dubail}. We can
also use evolution in imaginary time to predict the population of
low energy states following such a quench. In this case, instead
of a twist operator, we have a slit in the world sheet, whose
scaling dimension in a CFT was computed in \cite{CP} to be $c/16$.
Thus, in an infinite system, we have $P(E)dE\propto
E^{(c/8)-1}dE$.

Although the purpose of this article has been to show only that it
is possible \em in principle \em to measure R\'enyi entropies, it
is interesting to consider whether this proposal is at all
practical. As mentioned earlier, imperfections in the faithfulness
of the copies or of the precise details of the quench should be
unimportant as long as they occur at energy scales much less than
the gap $\Delta$ or the energy $E$ above the ground state. In
principle, $P(E)$ can be measured if the system couples weakly, in
a known manner, to other degrees of freedom whose energy
distribution can be spectrally analyzed. For example, if the
excited states of $H'$ can decay through a coupling to the
electromagnetic field, $P(E)$ should be recoverable from the
photon spectrum. In optical lattices, the energy $E$ could be
transferred to kinetic energy of the atoms, whose spectrum could
then be analyzed after removal of the trap. In both cases, in
order to produce an observable spectrum, it would of course be
necessary to consider a small but finite density of twists,
produced continuously at a low rate. Under such conditions, the
above analysis shows that we should obtain simple universal
behavior, characteristic of independent twists, for energies $E$
larger than mean density (in units where $v=1$), but much smaller
than the band width.

Although we have restricted attention to case of one dimension,
where twists are local, the analysis in principle extends to
higher dimensions. In particular it can be seen that the action of
twist operators can change the topology and so reveal the
entanglement entropy of topological phases. We have throughout
assumed that the ground state $|0\rangle$ is unique, and there are
interesting consequences, even in one dimension, when this is not
the case.

\acknowledgements The author thanks P.~Calabrese, B.~Doyon,
F.~Essler, P.~Fendley, E.~Fradkin and A.~Silva for useful
comments, and the KITP, Santa Barbara, for its hospitality. This
research was supported in part by the EPSRC under Grant
EP/D050952/1 and by the National Science Foundation under Grant
NSF PHY05-51164.


\begin{thebibliography}{99}
%
\bibitem{arealaw}  M.~Srednicki,  Phys. Rev. Lett. {\bf 71}, 666
(1993).
%
\bibitem{ccreview} For a review, see P.~Calabrese and J.~Cardy, J. Phys. A {\bf
42}, 504005 (2009).
%
\bibitem{holzhey} C.~Holzhey, F.~Larsen, and F.~Wilczek,  Nucl. Phys. B {\bf 424}, 44
(1994); C.~Callan and F.~Wilczek, Phys. Lett. B {\bf 333}, 55
(1994).
%
\bibitem{cc1} P.~Calabrese and J.~Cardy, J. Stat. Mech., 0406:P06002
(2004).
%
\bibitem{cct1}  P.~Calabrese, J.~Cardy and E.~Tonni, J. Stat. Mech., P11001
(2009).
%
\bibitem{cct2} P.~Calabrese, J.~Cardy and E.~Tonni,
J. Stat. Mech. P01021, (2011).
%
\bibitem{cirac} J.I.~Cirac and F.~Verstraete,
J. Phys. A: Math. Theor. {\bf 42}, 504004 (2009).
%
\bibitem{HE} P.~Horodecki and A.~Ekert, Phys. Rev. Lett. {\bf 89},
127902 (2002).
%
\bibitem{peschel} I.~Peschel, J. Phys. A {\bf 36}, L205 (2003).
%
\bibitem{hoyos} J.A.~Hoyos, A.P.~Vieira, N.~Laflorencie and
E.~Miranda, Phys. Rev. B {\bf 76}, 174425 (2007).
%
\bibitem{KRS}  I.~Klich, G.~Refael and A.~Silva,
Phys. Rev. A {\bf 74}, 032306 (2006).
%
\bibitem{kl} I.~Klich and L.~Levitov, Phys. Rev. Lett. {\bf 102}, 100502
(2009).
%
\bibitem{song1} H.F.~Song, C.~Flindt, S.~Rachel, I.~Klich, and K.~Le
Hur, arXiv:1008.5191.
%
\bibitem{song2} H.F.~Song, S.~Rachel, and K.~Le
Hur, arXiv:1002.0825.
%
\bibitem{fradkin} B.~Hsu, E.~Grosfeld and E.~Fradkin,
Phys. Rev. B {\bf 80}, 235412 (2009).
%
\bibitem{local} V. Eisler and I. Peschel, J. Stat. Mech. P06005
(2007).
%
\bibitem{cclocal} P.~Calabrese and J.~Cardy, J. Stat. Mech. P10004 (2007).
%
\bibitem{fidelity} L.~Campos Venuti and P.~Zanardi, Phys. Rev. Lett. {\bf 99},
096701 (2007).
%
\bibitem{dubail} J.~Dubail and J.-M.~St\'ephan, arXiv:1010.3716.
%
\bibitem{Pol} C.~De Grandi, V.~Gritsev and A.~Polkovnikov, Phys.
Rev. B {\bf 81}, 012303 (2009).
%
\bibitem{Silva} A.~Silva, Phys. Rev. Lett. {\bf 101}, 120603
(2008).
%
\bibitem{hastings} M.B.~Hastings, I.~Gonzalez, A.B.~Kallin and R.G.~Melko,
Phys. Rev. Lett. {\bf 104}, 157201 (2010).
%
\bibitem{Z1} P.~Zanardi, C.~Zolka and L.~Faoro, Phys. Rev. A {\bf
62}, 030301 (2000).
%
\bibitem{CAD} O.A.~Castro-Alvaredo and B.~Doyon, J.Stat.Mech. 1102:P02001,
(2011).
%
\bibitem{twist} L.J.~Dixon, D.~Friedan, E.J.~Martinec and S.H.~Shenker,
Nucl. Phys. B {\bf 282}, 13 (1987).
%
\bibitem{Xray} P.W.~Anderson, Phys. Rev. Lett. {\bf 18}, 1049
(1967).
%
\bibitem{Xraypower} K.D.~Schotte KD and U.~Schotte, Phys. Rev. {\bf 182},
479 (1969); I.~Affleck and A.W.W~Ludwig, J. Phys. A: Math. Gen.
{\bf 27}, 5375 (1994).
%
\bibitem{doyon1} J.L.~Cardy, O.A.~Castro-Alvaredo and B.~Doyon,
J. Stat. Phys. {\bf 130}, 129 (2007); O.A.~Castro-Alvaredo and
B.~Doyon, J. Phys. A {\bf 42}, 504006 (2009).
%
%
%
\bibitem{CP} J.~Cardy and I.~Peschel, Nucl. Phys. B {\bf 300}, 377, (1988).
%
\end{thebibliography}
\end{document}